\documentclass[aps,prb,showpacs,citeautoscript,twocolumn]{revtex4}

\usepackage{times}
\usepackage{bm}
\usepackage{graphicx}

\newcommand{\smeq}{\! = \!}

\newcommand{\smmi}{\! - \!}

\newcommand{\kf}{k_{\text{F}}}
\newcommand{\Ef}{E_{\text{F}}}

\newcommand{\be}{\begin{equation}}
\newcommand{\ee}{\end{equation}}
\newcommand{\bea}{\begin{eqnarray}}
\newcommand{\eea}{\end{eqnarray}}
\newcommand{\Ha}{{\hat H}}

\newcommand{\ci}{\mathrm{i}}

\begin{document}

\title{Design and control of spin gates in two quantum dots arrays}
\author{Gonzalo Usaj}
\affiliation{Instituto Balseiro and Centro At\'{o}mico Bariloche, Comisi\'{o}n Nacional de Energ\'{\i}a At\'{o}mica, (8400) San Carlos de
Bariloche, Argentina.}
\author{C. A. Balseiro}
\affiliation{Instituto Balseiro and Centro At\'{o}mico Bariloche, Comisi\'{o}n Nacional de Energ\'{\i}a At\'{o}mica, (8400) San Carlos de Bariloche, Argentina.}
\begin{abstract}
We study the spin-spin interaction between quantum dots coupled through a two dimensional electron gas with spin-orbit interaction. We show that the interplay between transverse electron focusing and spin-orbit
coupling allows to dynamically change the \textit{symmetry} of the effective spin-spin Hamiltonian. That is, the interaction can be changed from Ising-like to Heisenberg-like and vice versa. The sign and magnitude of the coupling constant can also be tuned. 
\end{abstract}
\date{\today}

\maketitle

Coherent control and measurement of quantum spins are at the heart
of new technologies with great potential value for information
processing \cite{Spintronicsbook,QCbook}. 
This has lead to a great activity in the field of quantum spin control in
solid state devices \cite{LossD98,BurkardLD99,HuS00,DivicenzoBKBW00,PotokFMU02,Elzerman03,KatoMGA04nature,KatoMGA05}.
Since the seminal work by Loss and DiVincenzo \cite{LossD98}, the
exchange gate is the central tool underlying most of the proposals
for spin manipulation in solid state devices based on quantum dots
(QDs) \cite{BurkardLD99,HuS00,DivicenzoBKBW00}.  
 The exchange gate is founded on the Heisenberg interaction
between localized spins  and so far nearly all
implementations for such control are based on an `on/off' setup---the interaction is either active or
inactive. Furthermore, when the exchange gate is controlled
by electrical gates, the control implies to `open' or `close'
the QDs, changing their coupling to the environment, their shape and
thus their detailed internal electronic structure. The question then is
: Is it possible to engineer a predefined
spin-spin interaction between QDs and then change its magnitude,
sign and symmetry with a negligible impact on the internal structure
of the dots?

In a recent work, we analyzed a way to tune the amplitude and sign
of the spin coupling \cite{UsajLB05}. Here we go a step forward and show
 how to design a Heisenberg or an Ising-like interaction of the desired
magnitude and sign of the coupling constant and then dynamically change one into the other by controlling a small
magnetic field---the control mechanism relies on the interplay between transverse electron focusing and spin-orbit
coupling \cite{vanHouten89,BeenakkerH91,UsajB04_focusing,RokhinsonLGPW04}. 
This opens up the possibility to manipulate spin-spin 
Hamiltonians in solid state devices as it is done today with NMR techniques in molecules \cite{Ernstbook}.

The setup consists of two QDs at the edges of two electron gases as
schematically shown in Fig. 1a, with an interdot distance $d$ of the order of $1\mu$m.
Present semiconducting
heterostructure technology allows tailoring this structure in two
dimensional electron gases (2DEG). In the Coulomb blockade regime,
the QDs can be gated to have an odd number of electrons so that they 
behave as magnetic objects. In what follows we describe them as
localized $\frac{1}{2}$ spins.  The virtual tunneling of electrons
between the dots and the 2DEG leads to a Kondo coupling between the
localized spins $\vec{\bm{S}}_{i}$ and the 2DEG spins described by
the following Hamiltonian \cite{SchriefferW66}:
\begin{equation}
\Ha_{K}\!=\!\sum_{i,\eta ,\eta ^{\prime }}J_{i}\vec{\bm{S}}%
_{i}\cdot \psi _{\eta \sigma }^{\dagger }(R_{i})\frac{\vec{%
\bm{\sigma }}_{\sigma \sigma ^{\prime }}}{2}\psi _{\eta ^{\prime
}\sigma ^{\prime }}^{}(R_{i}) \label{Hk}
\end{equation}
where $i\smeq1,2$ indicates the left and right QD respectively, $\psi
_{\eta \sigma }^{\dagger }(R_{i})$ creates an electron with spin
$\sigma $ in a Wannier-like orbital centered around the coordinate
$R_{i}$ of the $i$-th QD at the upper ($\eta \!=\!1$) or lower
($\eta \!=\!2$) plane. The spacial extension of the Wannier orbital
depends on the opening of the QDs. This coupling leads to a RKKY-like interaction between the QDs
spins that takes the general form \cite{ImamuraBU04}:
\begin{equation}
\Ha_{J}\smeq-\frac{J_{1}J_{2}}{4\pi }\mathrm{Im}\!\!\int\!\! d\omega f(\omega )\mathrm{Tr}\left(%
\vec{\bm{S}}_{1}\cdot \vec{\bm{\sigma }}\,%
\bm{G}(1,2)\vec{\bm{S}}_{2}\cdot \vec{\bm{%
\sigma }}\,\bm{G}(2,1)\right)
\label{HJ}
\end{equation}
where $f(\omega )$ is the Fermi function and the $2\!\times\!2$ matrix $\bm{%
G}(i,j)$ is the Fourier transform of the retarded electron
propagator whose elements are $G_{\sigma \sigma ^{\prime
}}(i,j,t\smmi t^{\prime })\smeq-\ci\theta (t\smmi t^{\prime })\times
$ $\sum_{\eta ,\eta ^{\prime }}\left\langle \{\psi _{\eta \sigma
}(R_{i},t),\psi _{\eta ^{\prime }\sigma ^{\prime }}^{\dagger
}(R_{j},t^{\prime })\}\right\rangle $.
When the electron's spin is conserved along the electron propagation
between QDs, $\bm{G}(i,j)$ is diagonal in the spin index and the
spin-spin Hamiltonian (\ref{HJ}) reduces to the Heisenberg one $\Ha_{J}\smeq J\vec{\bm{S}}%
_{1}\cdot \vec{\bm{S}}_{2}$ with the usual RKKY-like exchange
$J\smeq J_{1}J_{2}/2\pi\, \mathrm{Im}\int\! d\omega f(\omega )G_{\uparrow
\uparrow }(1,2)G_{\downarrow \downarrow }(2,1)\,$.
\begin{figure}[t]
   \begin{center}
   \includegraphics[height=2.3cm]{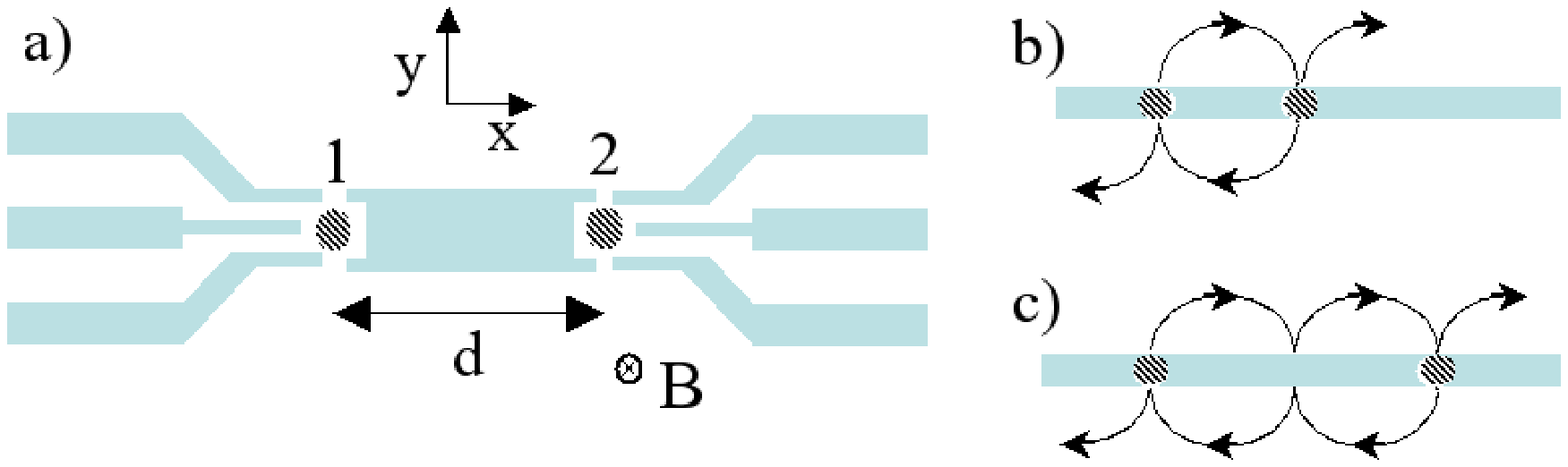}\\
   \includegraphics[height=3.cm]{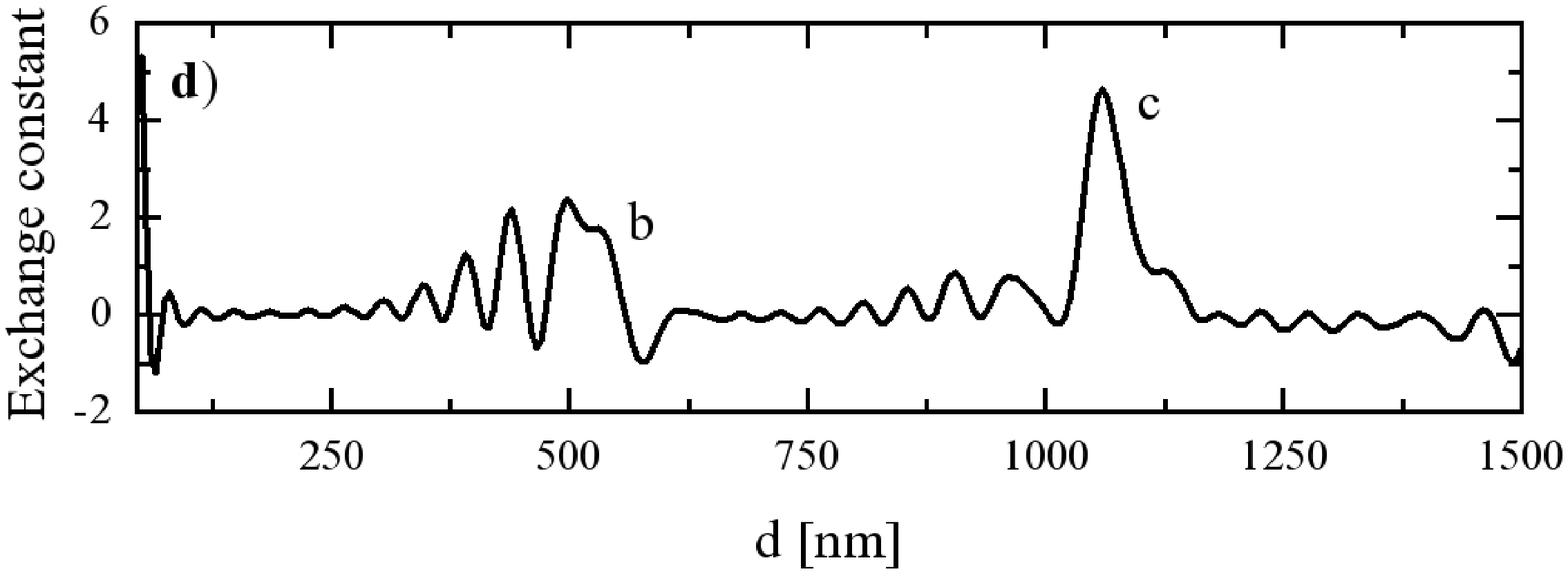}
   \end{center}
   \caption{\textbf{a}) Schematic view of the two QDs device. 
   \textbf{b}) and \textbf{c}) illustrate the semiclassical orbits
corresponding to the first and second focusing conditions, respectively.
Arrows indicate the direction of motion.
\textbf{d}) Exchange coupling constant $J$ as a function of the interdot distance. A large value of $J$ around the first (b)
and second (c) focusing condition is clearly seen. The
parameters correspond to a 2DEG with $m^{\ast
}\smeq0.067 m_{e}$ and $\Ef\smeq5$meV. The magnetic field
is $B_{z}\smeq227$mT and the temperature is set to zero. The exchange constant is normalized to its value at $d\!\approx\!\lambda_F$, for $B_z\smeq0$ and without SO.}
   \label{esquema}
\end{figure}

The presence of a small magnetic field $B_{z}$ perpendicular to the 2DEG creates edge
states that dominate the electron scattering from objects placed at the 2DEG
edges. The interaction between QDs is then mediated by these edge states
and the propagators are mainly due to the semiclassical orbits shown in
Figs. 1b and 1c. Due to the chiral nature of
these orbits, the intra-plane scattering,
described by the terms in Eq. (\ref{Hk}) with $\eta \smeq\eta ^{\prime }$, give
forward scattering while the inter-plane terms (with $\eta \neq \eta
^{\prime }$) describe the backward scattering.  Only the inter-plane backward scattering processes
contribute to the effective interaction. In other words, each propagators in
Eq. (2) is due to contributions from only one plane. As the external field
increases the cyclotron radii of these orbits decrease: $r_{c}\smeq\hbar
kc/eB_{z}$ with $k$ the electron wavevector. The focusing fields are those for
which the interdot distance $d$ is commensurate with the cyclotron radius $%
r_{c}$ of electrons at the Fermi energy ($\Ef$), that is $d\smeq2nr_{c}\smeq2n$ $\hbar
\kf c/eB_{z}$ with $n$ an integer number. At the focusing fields, the
electrons at the Fermi level scattered by one QD are focused onto the other
leading to an amplification of the exchange integral $J$ . The numerical
result is shown in Fig. 1d where, for the sake of comparison with the
conventional RKKY interaction, the exchange integral $J$ is plotted as a
function of the interdot distance for a fixed magnetic field. These results
were obtained using a finite differences technique \cite{UsajB04_focusing} for a system with
an effective electronic mass $m^{\ast }\smeq0.067m_{e}$ and $\Ef\smeq5$meV,  
corresponding to an electron density of $1.5\times 10^{11}/cm^{2}
$. With these parameters, the focusing amplification of the exchange
integral is clearly observed. In the semiclassical picture, the first
focusing condition ($n\smeq1$) corresponds to a direct propagation of the
electrons from one QD to the other; in the second one ($n\smeq2$) the electron
bounces once at the 2DEG edge. For interdot distances of the order of $1\mu
$m, the magnetic fields for the first focusing conditions ($n\smeq1$ or $2$) is
small and neglecting the Zeeman spin splitting due to the external field is
a good approximation. It is worth mentioning that, in similar geometries,
the electron focusing due to small magnetic fields is clearly observed in
transport experiments \cite{vanHouten89,PotokFMU02,FolkPMU03}. 

\begin{figure}[t]
   \centering
   \includegraphics[height=1.65cm]{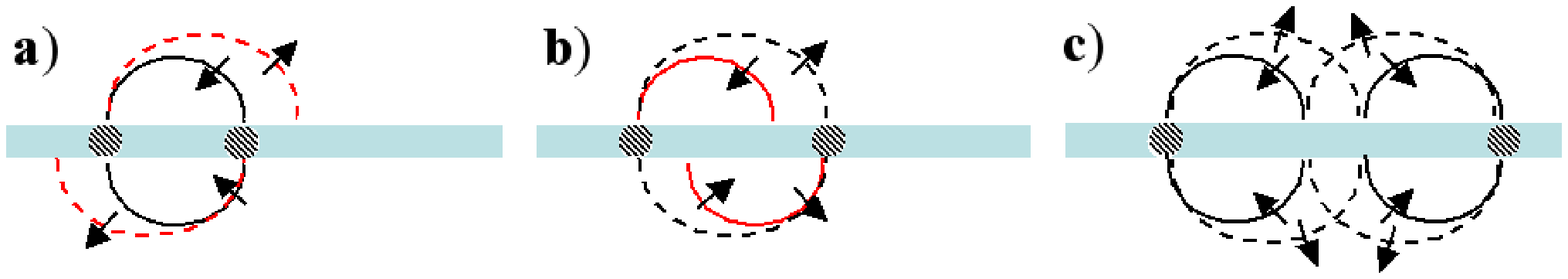}\\
   \includegraphics[height=3.4cm]{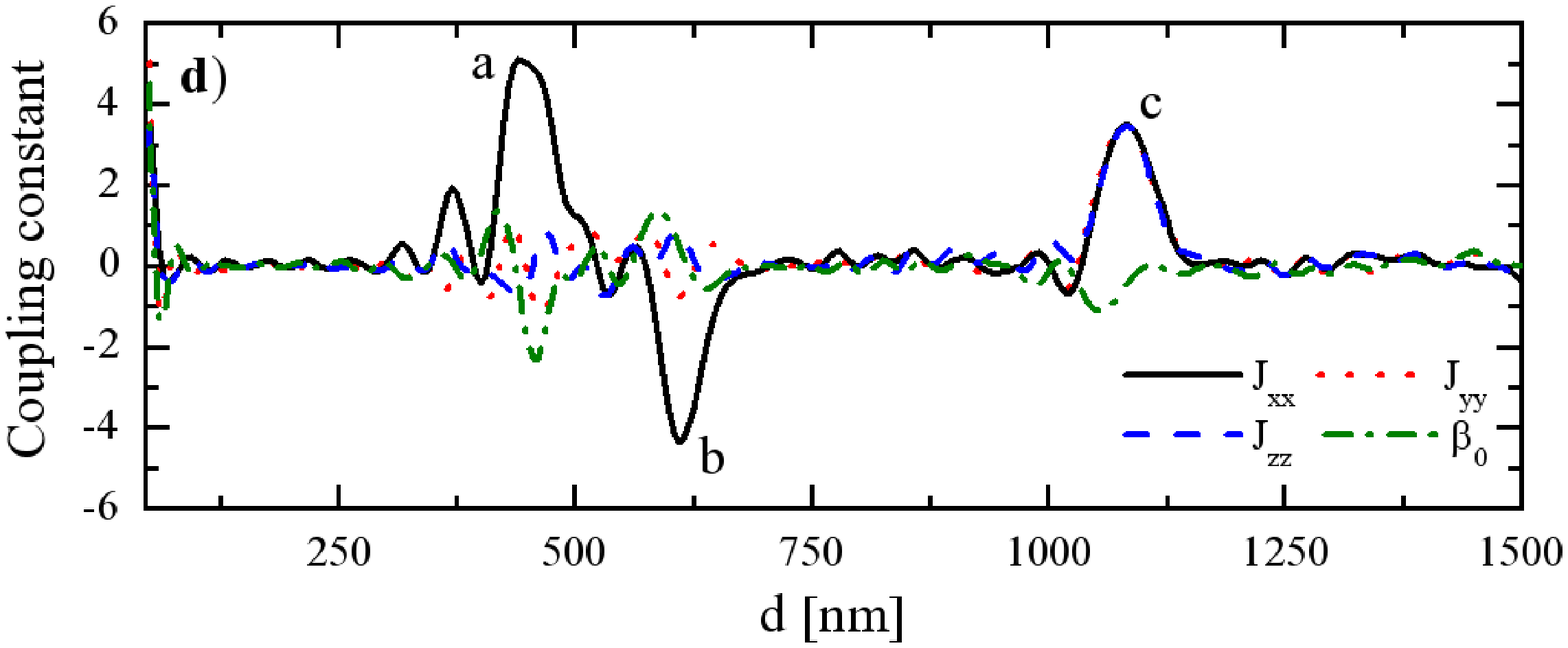}
   \caption{Schematic representation of the first focusing condition for the
smallest (a) and largest (b) cyclotron radius and of the second focusing condition (c) in
the presence of SO coupling. The arrows indicate the spin orientation. In
(d) the four coupling constants of Hamiltonian (\ref{Hgen}) are shown as a function of $d$. The SO parameter is taken to be $\alpha \smeq15$meVnm, and the other parameters as in Fig. 1. }
   \label{esquema_SO}
\end{figure}

In systems with strong spin-orbit (SO) coupling, new effects arise. We
consider a Rashba SO interaction in the 2DEG \cite{Rashba60,BychkovR84}. This interaction is due to the
inversion asymmetry of the confining potential and it is described by the
Hamiltonian $H_{SO}\smeq\alpha /\hbar (p_{y}\sigma _{x}-p_{x}\sigma _{y})$ where
$p_{\gamma }$ are the components of the canonical momentum of the 2DEG
electrons and $\sigma _{\gamma }$ the spin operators. The SO coupling acts
as a strong in-plane magnetic field proportional to the momentum. This
breaks the spin degeneracy leading to two different conduction bands \cite{Rashba60}. 

In the
presence of a small magnetic field perpendicular to the gas plane, each band
leads to a different cyclotron radius. These two radii manifest as two
distinct focusing fields for the first ($n\smeq1$) focusing condition \cite{UsajB04_focusing}. This
splitting has been observed by Rokhinson \textit{et al.} \cite{RokhinsonLGPW04} in a p-doped
GaAs/AlGaAs heterostructure. The spin texture of the orbits is such that, for
small fields (large cyclotron radii), the electron's spin adiabatically rotates
along the semiclassical orbit, being perpendicular to the momentum, as schematically shown in Fig. 2. 
In order to describe the magnetic scattering of electrons in
this case, it is convenient to quantize the spin along the
$x$-axis. Then, around the first focusing condition the propagators $%
G_{+,-}(i,j)$ and $G_{-,+}(i,j)$ dominate the interdot coupling, here the
spin index $\pm $ indicate the two spin projection. The
interdot interaction is then approximately given by an Ising term $%
\Ha_{I}\smeq J_{xx}S_{1x}S_{2x}$ with coupling constant given by
\begin{equation}
J_{xx}\smeq\frac{J_{1}J_{2}}{4\pi }\mathrm{Im}\int d\omega f(\omega
)G_{+,-}(i,j)G_{-,+}(j,i)
\end{equation}
where $i\smeq1$ and $j\smeq2$ or $i\smeq2$ and $j\smeq1$ depending on which cyclotron radius
contributes to the focusing. This result can be visualized in terms of the
semiclassical trajectories shown in Figs. 2a and 2b: for a SO
coupling strong enough to split the focusing condition, the inter-plane spin-flip backscattering 
mixes the two cyclotron radii living the electron out of the
focusing condition. Thus, these spin-flip processes can not contribute to the
coupling. The interdot interaction is then due to non-spin flip processes
of electrons that are back-scattered. This defines the symmetry axis of the resulting Ising interaction.

\begin{figure}[t]
   \centering
   \includegraphics[width=0.47\textwidth]{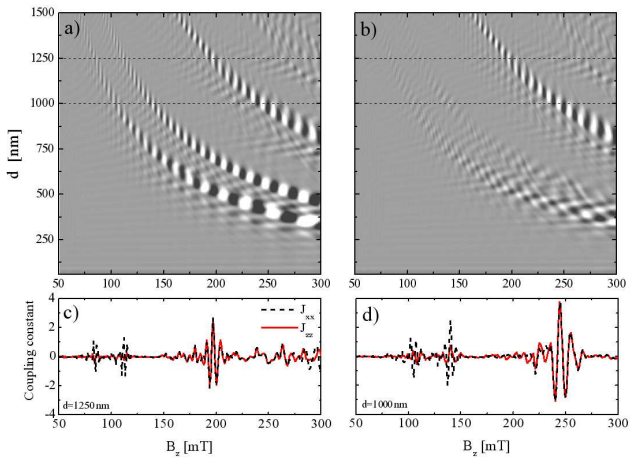}
   \caption{The exchange constants $J_{xx}$ (a) and $J_{zz}$ (b) as a function
of both the magnetic field $B_{z}$ and the interdot distance $d$. Black (white) corresponds to large negative (positive) values of the exchange
constants. Along the hyperbolas defined by the focusing
conditions large amplitude oscillations are observed. The first focusing
condition is split as explained in the text. (c) and (d) show slices taken
at $d\smeq1\mu$m in (c) and $d\smeq1.25\mu$m in (d); $%
J_{xx}$ and $J_{zz}$ are shown with dashed and solid lines
respectively. The isotropic nature of the exchange coupling
at the second focusing condition is apparent. Parameters as in Figs. 1 and 2.}
   \label{mapa}
\end{figure}

At the second focusing condition, the system operates in a different way
(see Fig. 2c). There are two important effects to consider: \textit{i})
the orbits with different cyclotron radii are mixed at the bouncing point due to spin conservation, 
and \textit{ii}) along the trajectories from one
QD to the other the electron's spin completes a $2\pi $ rotation. As a
consequence, the two orbits contribute to the exchange integral and $G_{+,+}(i,j)$
and $G_{-,-}(i,j)$ dominate the spin-spin coupling. In this way the 
rotational symmetric Heisenberg coupling is recovered.

For arbitrary external field, Hamiltonian (\ref{HJ}) can be written as fully
anisotropic Heisenberg model plus a Dzyaloshinski-Moriya term 
\begin{equation}
{\hat{H}}_{J}\smeq\sum_{\gamma }J_{\gamma \gamma }\,S_{1\gamma}S_{2\gamma }+%
\vec{\beta}\cdot \left( \vec{\bm{S}}%
_{1}\times \vec{\bm{S}}_{2}\right)
\label{Hgen}
\end{equation}
where $\vec{\beta}\smeq(0,\beta_{0},0)$.
Hamiltonian (\ref{Hgen}) is a particular case of a more general Hamiltonian
including SO effects \cite{Dzyaloshinski58,Moriya60,BonesteelSD01}. In
our case, due to the symmetry of our geometry, there are only four independent parameters: $J_{\gamma
\gamma }$ with $\gamma \smeq x,y,z$ and $\beta_{0}$. The
numerical results for these coupling constants are shown in Fig.
2d. As argued above, around the first focusing condition the system
behaves as an Ising like model: the dominant coupling $J_{xx}$ shows a large amplification when the interdot
distance matches each one the two cyclotron orbits. The relative
amplitude and sign of $J_{xx}$ in these peaks depends on both the
external field and the Fermi energy. At the second focusing
condition the system behaves as an isotropic Heisenberg model ($J_{xx}\smeq J_{yy}\smeq J_{zz}$) with a
small anisotropic correction  ($|\beta_{0}/J_{xx}|\ll 1$).
Figures 3a and 3b show the dominant couplings, $J_{xx}$ and $J_{zz}$ respectively, as a function of 
$B_{z}$ and $d$. The magnetic field not only
can turn on and off each coupling but a fine tune around the focusing fields can
change their sign too (see Figs. 3c and 3d) \cite{UsajLB05}.

There is a variety of systems that are potentially appropriate to observe
these effects. While in n-doped GaAs/AlGaAs heterostructures the spin orbit is
small, systems like p-doped GaAs/AlGaAs or InGaAs heterostructures present a
large SO coupling. The nature of the SO effect depends on the system.
Effects like the ones discussed in this work are also present in systems
with Dresselhaus SO coupling. Furthermore, the external control of the relative magnitude of both contributions
to the SO coupling \cite{MillerZMLGCG03}, could allow the control the quantization axis of the Ising-like interaction.

In summary, we have shown that the interplay between transverse electron focusing and
spin orbit interaction gives a unique opportunity to control and tune the
spin-spin interaction between QDs without inducing big changes in their internal structure. 
When the SO coupling is large, it leads to a spin-dependent focusing condition (for $n=1$), resulting in a highly anisotropic Ising-like interaction. However, by doubling the external field a
Heisenberg gate with a small correction of the Dzyaloshinski-Moriya type is
recovered. In the context of quantum computing, there are strategies to
eliminate or control these small corrections to the Heisenberg gate \cite{BonesteelSD01,StepanenkoBDGL03,StepanenkoB04}.
The proposed setup can be extended to three or more QDs in a linear array. An
array with different interdot distances and with extra gates used to
blockade the focusing may be used to independently control the interdot
interactions.

This work was partially supported by ANPCyT
Grants No 13829 and 13476 and Fundaci\'on Antorchas, Grant 14169/21. GU acknowledge 
support from CONICET.

\end{document}